\documentclass[12pt]{article}
\usepackage{amssymb}
\begin{document}
\title{\bf Exploring Plane Symmetric Solutions in $f(R)$ Gravity}

\author{M. Farasat Shamir\thanks{farasat.shamir@nu.edu.pk}\\\\
Department of Sciences and Humanities, \\National University of
Computer and Emerging Sciences,\\ Lahore Campus, Pakistan.}

\date{}

\maketitle
\begin{abstract}
The modified theories of gravity, especially the $f(R)$ gravity, have attracted much attention in the last decade.
This paper is devoted to exploring plane symmetric solutions
in the context of metric $f(R)$ gravity. We
extend the work on static plane symmetric vacuum solutions in $f(R)$
gravity already available in literature \cite{farasat, farasats}.
The modified field equations are solved using the
assumption of both constant and non-constant scalar curvature.
Some well known solutions have been recovered with power law and
logarithmic forms of $f(R)$ models.
\end{abstract}


\section{Introduction}

Astrophysical data coming from different sources such as Cosmic Microwave
Background fluctuations \cite{CMB}, Supernovae Type Ia experiments \cite{SN},
X-ray experiments \cite{xray} and large scale structure
\cite{LSS} have revealed a completely different picture of our universe.
All these observations suggest that the universe is expanding with an
accelerated rate. The phenomenon of dark energy and
dark matter is another topic of discussion \cite{de}. It
was Einstein who first gave the concept of dark energy and
introduced the small positive cosmological constant in the field equations. But after sometime,
he remarked it as the biggest mistake in his life. However, it is
now believed that our universe is filled with exotic cosmic fluid known as dark energy with strong
negative pressure and the cosmological constant may be a suitable
candidate for dark energy. There exist two basic models for dark energy. In the first model it is associated with empty space
and remains constant throughout the spacetime suggesting the need of cosmological constant in the field equations.
Second model proposes that it varies over the spacetime and cosmic expansion is achieved by a scalar field.
Different models have been proposed involving scalar field, i.e. quintessence \cite{QUINT},
k-essence \cite{KESSENCE}, chaplygin gas \cite{Chaplygin} and phantom models \cite{PHANTOM}.
It has been predicted that $96\%$ energy of the universe is either dark
energy or dark matter ($76\%$ dark energy and $20\%$ dark matter) \cite{de}.
Matter and energy domination seems to be a justified reason for this accelerating phase.
We can describe the dark energy with an Equation of State (EoS) parameter $\omega=p/\rho$,
where $\rho$ and $p$ represent the energy density and pressure of dark energy respectively.
It has been established that the expansion of the universe is accelerating when $w\approx-1$ \cite{nature}.
The universe is found to have quintessence dark era when $\omega>-1$ while the phantom like dark energy exists in the region where $\omega<-1$. The universe with phantom like dark energy ends up with a finite time future singularity known as big rip or cosmic doomsday \cite{DOOMSDAY}. Some other observations like rotational velocities of
galaxies, the temperature distribution of hot gas in galaxies, gravitational lensing of background objects by galaxy clusters and the observed fluctuations in the cosmic microwave background radiation have indicated
the presence of additional gravity, which may be justified by the existence of dark matter in the universe. According to some astrophysicists, modified theories of gravity may explain this phenomenon of dark matter and dark energy which seem to responsible for current cosmic expansion.

Nowadays, an extended theory known as $f(R)$ theory of gravity has attracted much attention of the researchers.
It is believed that modification of Einstein's theory
with some inverse curvature terms may cause an increase in gravity
which justifies accelerated expansion \cite{invR}. However, modified gravity is known to be unstable
with inverse curvature terms and does not pass some solar system tests
\cite{unstable}. This discrepancy can be addressed by using higher
derivative terms. Moreover, squared curvature terms can be used to achieve viability \cite{16}.
It is now expected that the
cosmic expansion can be explained if some suitable powers of
curvature are included in the Einstein-Hilbert action.
The dark matter problems can also be
addressed using viable $f(R)$ gravity models \cite{V1}.
Thus it would be interesting to investigate modified or alternative theories of gravity. The $f(R)$
theory of gravity, which involves a generic function of Ricci scalar
in standard Einstein-Hilbert lagrangian, is an attractive choice.

Recent literature \cite{25}-\cite{8} shows keen interest in exploring
different issues in modified $f(R)$ theories of gravity.
Spherically symmetric solutions are the most widely explored exact solutions in $f(R)$ gravity.
Multam$\ddot{a}$ki and Vilja \cite{20} investigated spherically symmetric vacuum solutions and
it was found that the set of the field equations in
$f(R)$ gravity provided the Schwarzschild deSitter metric. The
same authors \cite{21} found the perfect fluid solutions and
concluded that unique form of $f(R)$ was not obtained.
Spherically symmetric solutions in $f(R)$ gravity using the Noether symmetry have been explored by
Capozziello et al. \cite{22}. The exact static spherically symmetric spacetimes solutions in $f(R)$
gravity coupled to non-linear electrodynamics have been analyzed by Hollenstein and Lobo \cite{24}.
It seems interesting, at least from theoretical point of view, to
consider other exact solutions of the field equations in $f(R)$
gravity. Azadi et al. \cite{23} found cylindrically
symmetric vacuum solutions in this theory.
In another paper \cite{024}, Momeni gave more general results using cylindrical symmetry in  $f(R)$
theory.

Here we focus our attention to investigate the exact
solutions of static plane symmetric solutions in metric $f(R)$
gravity. In particular,
some solutions have been found using the assumption of both constant and non-constant scalar
curvature. Some well known solutions already available in general theory of relativity (GR) have been recovered
in the presence of important $f(R)$ gravity models. The paper is organized as follows: In section
\textbf{2}, we introduce the field equations
in the context of $f(R)$ gravity. Section \textbf{3} is used to
find plane symmetric solutions. In section \textbf{4}, we briefly discuss the physical importance of the solutions. The results are summarized and concluded in the last section.

\section{Some Basics of $f(R)$ Gravity }

Mainly two approaches exist in $f(R)$ theories of gravity. The
first is known as ``metric approach" in which the connection is the
Levi-Civita connection and the action is varied with
respect to the metric. The second approach is called the ``Platini
formalism" in which the connection and the metric are considered
independent of each other and the variation is done for the two
parameters independently. Here we use the metric
approach to explore the exact solutions.

The $f(R)$ theory of gravity is actually the generalization or modification of GR. The action for
$f(R)$ gravity is given by \cite{20}
\begin{equation}\label{1}
S=\int\sqrt{-g}(\frac{1}{16\pi{G}}f(R)+L_{m})d^4x.
\end{equation}
Here $f(R)$ is a generic function of the Ricci scalar and $L_{m}$ is known as
the matter Lagrangian. It may be observed that this action is
obtained by just replacing $R$ with $f(R)$ in the standard
Einstein-Hilbert action. The corresponding field equations are found
by varying the action with respect to the metric $g_{\mu\nu}$
\begin{equation}\label{2}
F(R)R_{\mu\nu}-\frac{1}{2}f(R)g_{\mu\nu}-\nabla_{\mu}
\nabla_{\nu}F(R)+g_{\mu\nu}\Box F(R)=\kappa T_{\mu\nu},
\end{equation}
where $T_{\mu\nu}$ is the standard matter energy-momentum tensor and
\begin{equation}\label{3}
F(R)\equiv df(R)/dR,\quad\Box\equiv\nabla^{\mu}\nabla_{\mu}
\end{equation}
with $\nabla_{\mu}$ as the covariant derivative. The Eqs.(\ref{2}) are the fourth order partial differential equations in
the metric tensor. These
equations may reduce to the field equations of GR if we take $f(R)=R$.\\\\
Now contraction of the field equations yields
\begin{equation}\label{4}
F(R)R-2f(R)+3\Box F(R)=\kappa T.
\end{equation}
For vacuum case, this reduces to
\begin{equation}\label{5}
F(R)R-2f(R)+3\Box F(R)=0.
\end{equation}
Eq.(\ref{5}) gives an important relationship between $F(R)$ and $f(R)$ which later on may be
used to simplify the field equations and to evaluate $f(R)$. It can be seen from Eq.(\ref{5}) that any metric having constant Ricci scalar
, say $R=R_{0}$, is a solution of the contracted equation
(\ref{5}) if the following equation holds
\begin{equation}\label{6}
F(R_{0})R_{0}-2f(R_{0})=0.
\end{equation}
This condition is known as ``constant curvature condition". Further, the
differentiation of Eq.(\ref{5}) with respect to $x$ gives
\begin{equation}\label{6234}
F'(R)R-R'F(R)+3(\Box F(R))'= 0.
\end{equation}
These conditions Eqs.(\ref{6},\ref{6234}) were first derived by Cognola et al. \cite{37}.

\section{Plane Symmetric Solutions}

In this section, we will find plane symmetric static solutions of the field
equations in metric $f(R)$ gravity. We first use the
constant scalar curvature $(R=constant)$ to find the solutions.
We will also take non-constant curvature condition to obtain
solutions of the static plane symmetric spacetimes  in $f(R)$ gravity.

\subsection{Plane Symmetric Spacetimes}

The general static plane symmetric spacetime is
\begin{equation}\label{32}
ds^{2}=A(x)dt^{2}-C(x)dx^{2}-B(x)(dy^{2}+dz^{2}).
\end{equation}
For the sake of simplicity, we assume $C(x)=1$ so that the above
spacetime becomes
\begin{equation}\label{33}
ds^{2}=A(x)dt^{2}-dx^{2}-B(x)(dy^{2}+dz^{2}).
\end{equation}
The Ricci scalar turns out to be
\begin{equation}\label{34}
R=\frac{1}{2}[\frac{2A''}{A}-(\frac{A'}{A})^2
+\frac{2A'B'}{AB}+\frac{4B''}{B}-(\frac{B'}{B})^2],
\end{equation}
where prime denotes derivative with respect to $x$. Eq.(\ref{4}) can be rearranged as
\begin{equation}\label{9}
f(R)=\frac{3\Box F(R)+F(R)R-\kappa T}{2}.
\end{equation}
Using this value of $f(R)$ in the field equations, it follows that
\begin{equation}\label{10}
\frac{F(R)R_{\mu\nu}-\nabla_{\mu}\nabla_{\nu}F(R)-\kappa
T_{\mu\nu}}{g_{\mu\nu}} =\frac{F(R)R-\Box F(R)-\kappa T}{4}.
\end{equation}
The dependance of metric (\ref{33}) on $x$ suggests that one can view
Eq.(\ref{10}) as the set of differential equations for $A,~B$, $F$, $\rho$ and $p$.
From Eq.(\ref{10}), we can see that the combination
\begin{equation}\label{11}
A_{\mu}\equiv\frac{F(R)R_{\mu\mu}-\nabla_{\mu}\nabla_{\mu}
F(R)-\kappa T_{\mu\nu}}{g_{\mu\mu}}
\end{equation}
is independent of the index $\mu$ and hence
$A_{\mu}-A_{\nu}=0$ for all $\mu$ and $\nu$.
So $A_{0}-A_{1}=0$ yields
\begin{equation}\label{35}
[\frac{A'B'}{AB}+(\frac{B'}{B})^2-\frac{2B''}{B}]F-2F''+\frac{A'}{A}F'-2\kappa(\rho+p)=0.
\end{equation}
Also, $A_{0}-A_{2}=0$ gives
\begin{equation}\label{36}
[\frac{A''}{A}-\frac{1}{2}(\frac{A'}{A})^2+\frac{A'B'}{2AB}-
\frac{B''}{B}]F+(\frac{A'}{A}-\frac{B'}{B})F'-2\kappa(\rho+p)=0.
\end{equation}
Thus we obtain two differential equations with five
unknowns namely $A,~B$, $F$, $\rho$ and $p$. These equations seem difficult to solve due to their highly nonlinear nature.
However, we investigate some
solutions using the assumptions of both constant and non-constant curvature.

\subsection{Solutions With Constant Curvature Assumption}

Here we consider constant curvature case, say $R=R_{0}$. Thus we have
\begin{equation}\label{15}
F'(R_{0})=0=F''(R_{0}).
\end{equation}
It is clear that any solution found for GR will be found for specific version of $f(R)$
theory. In particular, the constant curvature solutions found in $f(R)$ are already available solutions in GR.

\subsection*{Case I}

Using condition (\ref{15}), Eqs.(\ref{35},~\ref{36}) reduce to
\begin{equation}\label{00035}
[\frac{A'B'}{AB}+(\frac{B'}{B})^2-\frac{2B''}{B}]F_0-2\kappa(\rho+p)=0.
\end{equation}
\begin{equation}\label{00036}
[\frac{A''}{A}-\frac{1}{2}(\frac{A'}{A})^2+\frac{A'B'}{2AB}-
\frac{B''}{B}]F_0-2\kappa(\rho+p)=0.
\end{equation}
We can describe the dark energy with EoS parameter $\omega=p/\rho$,
where $\rho$ and $p$ represent the energy density and pressure of dark energy.
It has been established that the expansion of the universe is accelerating when $w\approx-1$ \cite{nature}.
In this case, Eq.(\ref{00035}) and Eq.(\ref{00036}) reduce to
\begin{eqnarray} \label{37}
\frac{A'B'}{AB}+(\frac{B'}{B})^2-\frac{2B''}{B}=0,\\\label{38}
\frac{A''}{A}-\frac{1}{2}(\frac{A'}{A})^2+\frac{A'B'}{2AB}-
\frac{B''}{B}=0.
\end{eqnarray}
These equations can be solved using the power law assumption, i.e.,
$A\propto x^{r}$ and $B\propto x^{l}$, where $r$ and $l$ are any
real numbers. Thus we use $A=k_1x^{r}$ and $B=k_2x^{l}$, where
$k_1$ and $k_2$ are constants of proportionality. It follows that
\begin{equation}\label{41}
\quad r=-\frac{2}{3}, \quad l=\frac{4}{3}
\end{equation}
and hence the solution becomes
\begin{equation}\label{42}
ds^{2}=k_1x^{-\frac{2}{3}}dt^{2}-dx^{2}-k_2x^{\frac{4}{3}}(dy^{2}+dz^{2}).
\end{equation}
These values of $r$ and $l$ lead to $R=0$.
This is the most basic possible solution and somehow trivial in constant curvature case. We
can re-define the parameters, i.e., $\sqrt{k_1}~t\longrightarrow
\tilde{t},~\sqrt{k_2}~y\longrightarrow \tilde{y}$ and
$\sqrt{k_2}~z\longrightarrow \tilde{z}$, so that the above metric takes the
form
\begin{equation}\label{42a}
ds^{2}=x^{-\frac{2}{3}}d\tilde{t}^{2}-dx^{2}-x^{\frac{4}{3}}(d\tilde{y}^{2}+d\tilde{z}^{2})
\end{equation}
which is the same as Taub's metric \cite{30}.

\subsection*{Case II}

Now we assume $B={A}^{n}$ so that the subtraction of
Eq.(\ref{00035}) and Eq.(\ref{00036}) gives
\begin{equation}\label{43}
(3n+1){A'}^2-2(n+1)AA''=0.
\end{equation}
This equation yields a solution given by
\begin{equation}\label{44}
A=k_3[(n-1)x+2k_4(n+1)]^{\frac{2(n+1)}{1-n}},
\end{equation}
where $k_3$ and $k_4$ are integration constants. Without loss of generality, we can choose
$k_3=1$ and $k_4=0$ so that Eq.(\ref{44}) takes the farm
\begin{equation}\label{46}
A=[(n-1)x]^{\frac{2(n+1)}{1-n}},
\end{equation}
and thus $B$ turns out to be
\begin{equation}\label{47}
B=[(n-1)x]^{\frac{2n(n+1)}{1-n}}
\end{equation}
and the solution metric takes the form
\begin{equation}
ds^{2}=[(n-1)x]^{\frac{2(n+1)}{1-n}}dt^{2}-dx^{2}-[(n-1)x]^{\frac{2n(n+1)}{1-n}}(dy^{2}+dz^{2}).
\end{equation}
It would be worthwhile to mention here that we can recover Taub's solution when $n=-2$.

\subsection{Solutions Without Constant Curvature Assumption}

Now we explore the solutions of modified  field equations without using the constant curvature assumption.
Subtracting Eq.(\ref{35}) and Eq.(\ref{36}), we obtain
\begin{equation}\label{51}
\bigg[\frac{A'B'}{AB}+\bigg(\frac{A'}{A}\bigg)^2+2\bigg(\frac{B'}{B}\bigg)^2-
2\bigg(\frac{A''}{A}+\frac{B''}{B}\bigg)\bigg]F+2\frac{B'}{B}F'-4F''=0.
\end{equation}
Due to highly non-linear nature of Eq.(\ref{51}), here we also use the assumption $B={A}^{n}$.
Thus the Eq.(\ref{51}) reduces to
\begin{equation}\label{52}
(3n+1)\bigg(\frac{A'}{A}\bigg)^2-2(n+1)\bigg(\frac{A''}{A}\bigg)+2n\frac{A'F'}{AF}-4\frac{F''}{F}=0
\end{equation}
and the Ricci scalar turns out to be
\begin{equation}\label{53}
R=\frac{1}{2}\bigg[(3n^{2}-2n-1)\bigg(\frac{A'}{A}\bigg)^2+(4n+2)\frac{A''}{A}\bigg].
\end{equation}

For this purpose, We follow the approach of Nojiri and
Odintsov \cite{Unified cosmic history in modified gravity from
F(R) theory to Lorentz non-invariant models} and take the
assumption $F(R)\propto f_0R^m$, where $f_0$ is an arbitrary
constant.  So using Eqs.(\ref{52},\ref{53}) and after some tedious calculations we obtain\newpage
\begin{eqnarray}\nonumber
&&[8m(1+4n-2n^2-12n^3+9n^4)+16m^2(1+4n-2n^2-12n^3+9n^4)
\\\nonumber&&-1-3n+6n^2+10n^3-21n^4+9n^5]{A'}^{6}+[-8m(5+20n+2n^2-36n^3+9n^4)
\\\nonumber&&-32m^2(2+8n-n^2-18n^3+9n^4)+2(3+11n-4n^2-22n^3+21n^4-9n^5)]
\\\nonumber&&{A'}^{4}{A''}A+[-8m(1+4n+4n^2-9n^4)+32m^2(1+4n+n^2-6n^3)
\\\nonumber&&+4n(1+4n+n^2-6n^3)]{A'}^{3}A'''A^2+[8m(8+32n+17n^2-30n^3-9n^4)
\\\nonumber&&+16m^2(4+16n+4n^2-24n^3+9n^4)-4(3+13n+10n^2-8n^3)]{A'}^{2}{A''}^{2}A^2
\\\nonumber&&+[16m(1+4n+n^2-6n^3)-32m^2(2+8n+5n^2-6n^3)-8n(1+4n+4n^2)]
\\\nonumber&&{A'}{A''}{A'''}A^3+16m(1+4n+4n^2){A''''}{A''}A^4-8m(1+4n+n^2-6n^3){A''''}{A'}^2A^3
\\\nonumber&&+[16m^2(1+4n+4n^2)-16m(1+4n+4n^2)]{A'''}^2A^4+[-16m(2+8n\\\label{54}
&&+5n^2-6n^3)+8(1+5n+8n^2+4n^3)]{A''}^3A^3=0.
\end{eqnarray}
Many solutions can be reconstructed using this equation. However we discuss only three cases here.

\subsection*{Case III}

In this case we try to recover the Taub's solution. For this purpose, we substitute $A=x^\frac{-2}{3}$ in Eq.(\ref{54}).
After some lengthy calculations, we obtain a constraint equation
\begin{equation}\label{55}
18m^2+9m-1+n=0.
\end{equation}
We can obtain $B=x^{\frac{4}{3}}$ for $n=-\frac{1}{2}$. Thus Eq.(\ref{55}) reduces to
\begin{equation}\label{56}
12m^2+6m-1=0.
\end{equation}
The roots of Eq.(\ref{56}) turn out to be
$m={\frac{-3\pm\sqrt{21}}{12}}$. Thus, we have
\begin{equation}\label{57}
F(R)=f_0R^{\frac{-3\pm\sqrt{21}}{12}}.
\end{equation}
After integration, we obtain
\begin{equation}\label{58}
f(R)=\hat{f}_0(R)^{\frac{9+\sqrt{21}}{12}}+k_5,~~~~~f(R)=\check{f}_0(R)^{\frac{9-\sqrt{21}}{12}}+k_6,
\end{equation}
where $\hat{f}_0=\frac{12f_0}{9+\sqrt{21}}$, $\check{f}_0=\frac{12f_0}{9-\sqrt{21}}$ and $k_5$, $k_6$ are
integration constants. It has been proved that the terms with positive
powers of the curvature support the inflationary epoch \cite{16}. The corresponding Ricci scalar becomes
\begin{equation}\label{60}
R=\frac{2}{3x^2}.
\end{equation}
Using first root $m={\frac{-3+\sqrt{21}}{12}}$, EoS parameter $\omega$  and Eqs.(\ref{35},~\ref{36}),
the energy density of the universe turns out to be
\begin{equation}
\rho=\frac{-f_0}{18\kappa(1+\omega)}\bigg[\frac{21(\frac{2}{3})^{\frac{-3+\sqrt{21}}{12}}}{x^{\frac{1+\sqrt{21}}{6}}}+
\frac{5(\frac{-3+\sqrt{21}}{3})(\frac{2}{3})^{\frac{-15+\sqrt{21}}{12}}}{x^{\frac{9+\sqrt{21}}{6}}}+
\frac{4}{3x^6}\bigg].
\end{equation}
We can choose the sign of $f_0$ depending upon the values of $\omega$ to get the positive energy density.
Similarly, we can find expression for energy density in the case of other root $m={\frac{-3-\sqrt{21}}{12}}$.

\subsection*{Case IV}

Here we take $A=\frac{1}{x}$ in Eq.(\ref{54}) to obtain a constraint equation
\begin{equation}\label{61}
16m^2+8m-3n+3=0.
\end{equation}
Using this equation, it follows that
\begin{equation}\label{62}
B=x^{-\frac{16m^2+8m+3}{3}}.
\end{equation}
The solution metric takes the farm
\begin{equation}\label{63}
ds^{2}={\frac{1}{x}}dt^{2}-dx^{2}-x^{-\frac{16m^2+8m+3}{3}}(dy^{2}+dz^{2}).
\end{equation}
The corresponding Ricci scalar becomes
\begin{equation}\label{64}
R=\frac{3(n^2+2n+1)}{2x^2}.
\end{equation}
We can construct different $f(R)$ models for different values of $m$ satisfying Eq.(\ref{61}).
However, an interesting logarithmic form of $f(R)$ models is obtained for $m=-1$
\begin{equation}
f(R)=f_0ln(R)+k_7,
\end{equation}
where $k_7$ is an integration constant. Such type of logarithmic form was first introduced by Nojiri and Odintsov \cite{2933}. In this case Ricci scalar becomes to $R=\frac{98}{3x^2}$ and the solution metric takes the form
\begin{equation}\label{6293}
ds^{2}={\frac{1}{x}}dt^{2}-dx^{2}-\frac{1}{x^{\frac{11}{3}}}(dy^{2}+dz^{2})
\end{equation}
and matter density turns out to be
\begin{equation}
\rho=\frac{-f_0}{441\kappa(1+\omega)}\bigg[93+\frac{470596}{x^6}\bigg].
\end{equation}
Similarly for $m=-2$, we obtain
\begin{equation}
f(R)=-f_0{R}^{-1}+k_8,
\end{equation}
where $k_8$ is an integration constant. This model is also cosmologically important as it has been
proved that negative power of curvature serves as an effective dark energy supporting the cosmic acceleration \cite{16}.
Obviously one can work out the Ricci scalar, energy density and the solution metric in this case.

\subsection*{Case V}

Here we consider $A=e^{x}$ in Eq.(\ref{54}).
In this case, we obtain a constraint equation
\begin{equation}\label{65}
(n-1)(3n^2+2n+1)^2=0,
\end{equation}
which does not involve parameter $m$. So this choice will yield a solution for any
$f(R)$ model in power law or logarithmic form. The roots of Eq.(\ref{55}) turn out to be
\begin{equation}\label{66}
n=1,\quad \frac{-3\pm i\sqrt{3}}{6}.
\end{equation}
We discard the imaginary roots and consider the real value of $n$ to get a physical solution.
In this case the Ricci scalar turns out to be non-zero constant, i.e., $R=3$. The energy density is also constant here and the solution metric becomes
\begin{equation}\label{67}
ds^{2}=e^{x}(d{t}^{2}-d{y}^{2}-d{z}^{2})-dx^{2}.
\end{equation}
This corresponds to the well-known anti-deSitter spacetime
in GR \cite{32}.\\\\

\subsection{Physical Importance of the Solutions}

The spacetime admitting three parameter group of motions of the Euclidean
plane is said to possess plane symmetry and is known as a plane symmetric spacetime.
Such spacetime possesses many properties equivalent to those of spherical
symmetry. The plane symmetric spacetime has been extensively investigated by many
researchers from various standpoints. Taub \cite{33}, Bondi \cite{34}, Bondi and Pirani-Robinson \cite{35}
defined and studied plane wave solutions. They considered the concept of group of motions of spacetime which
played a fundamental role in plane gravitational waves. It has been established that the spacetime Eq.(\ref{33})
admits the plane wave solutions of GR field equations \cite{36}.

In this study, we have explored plane symmetric solutions in the context of $f(R)$ gravity. This is actually an extension of already done work \cite{farasat,farasats} where the solutions are with vacuum and constant curvature case only. Here we do not relax the conditions and generalize the already obtained solutions. The non-vacuum plane symmetric solutions provide Taub's universe with a singularity at $x=0$ which suggests the presence of black hole. Another solution (\ref{63}) suggests that an object falling into a black hole approaches the singularity at $x=0$. However, non-singular solution is obtained in the shape of an anti-deSitter spacetime. An anti-deSitter space is a GR like spacetime, where in the absence of matter or energy, the curvature of spacetime is naturally hyperbolic.
From geometrical point of view, an anti-de Sitter space has a curvature analogous to a flat cloth sitting on a saddle, with a very slight curvature because it is so large. Thus it would correspond to a negative cosmological constant. Anti-deSitter space can also be thought as empty space having negative energy, which causes this spacetime to collapse at a greater rate. The existence of quantum-corrected deSitter space has been predicted as an outcome of a nontrivial solution for constant curvature $R_0$ in $f(R)$ gravity \cite{37}.
One may play with the parameters of the theory under consideration in such a way that the
deSitter space can provide a solution to the cosmological constant problem. Thus the physical relevance of the solutions is obvious.

\section{Summary and Conclusion}

In this paper, we focuss ourselves to explore the plane symmetric solutions
in $f(R)$ gravity. We have considered the metric version of
the theory to find the exact solutions of field equations.
We would like to point out that most of the work in $f(R)$ gravity
has been done for vacuum static cases with constant curvature condition.
It can be interesting to find the solutions for non-static and non-vacuum
cases without using constant curvature condition.
So as a first step, we investigate plane symmetric solutions with non-vacuum case.
To our knowledge, this is the first attempt to investigate non-vacuum plane symmetric solutions
in $f(R)$ gravity without using constant curvature assumption.
We can assume the function of Ricci scalar arbitrarily to solve the field equations
but this gives fourth order highly non-linear differential
equations. The assumption of constant curvature (may be zero or
non-zero) seems to be most suitable and we can get some
solutions for constant scalar curvature. We have found two solutions with
this assumption and recovered the well known Taub's solution.

The solutions without the assumption of constant scalar curvature provide some important $f(R)$ gravity models.
Mainly we have explored three solutions in this context. First solution gives the Taub's spacetime
with power law forms of $f(R)$ models having positive curvature. It would be worthwhile to mention here that the terms with positive
powers of the curvature support the inflationary epoch. Ricci scalar is non-constant in this case.
Second solution also yields non-constant curvature and two important $f(R)$ models have been constructed in this case.
First model is in logarithmic form while second corresponds to negative power of curvature. It would be worthwhile to mention here that
negative power of curvature serves as an effective dark energy supporting the current cosmic acceleration.
The third case yields a well known solution which corresponds to anti-deSitter spacetime. It provides an arbitrary
$f(R)$ model in power law or logarithmic form. The Ricci scalar in this case is non-zero constant.
We have discussed five cases in this paper. However, many other cases can also be explored and different cosmologically important $f(R)$ models can be reconstructed.\\\\
\textbf{Acknowledgement}\\\\ The author would like to acknowledge National University
of Computer and Emerging Sciences (NUCES) for
funding support through research reward programme.
The author is also thankful to the anonymous reviewer
for valuable comments and suggestions to improve the paper.

\end{document}